\documentclass[conference,a4paper]{APSIPA2019}
\usepackage{multirow}
\usepackage{amsmath,graphicx}
\usepackage{amsmath}
\usepackage[psamsfonts]{amssymb}
\usepackage{amsxtra}
\usepackage{threeparttable}

\begin{document}

\title{Utterance-level Permutation Invariant Training with Latency-controlled BLSTM for Single-channel Multi-talker Speech Separation}

\author{%
\authorblockN{%
Lu Huang\authorrefmark{1} and
Gaofeng Cheng\authorrefmark{2}\authorrefmark{3} and
Pengyuan Zhang\authorrefmark{2}\authorrefmark{3} and
Yi Yang\authorrefmark{1} and
Shumin Xu\authorrefmark{4} and
Jiasong Sun\authorrefmark{1}
}
\authorblockA{%
\authorrefmark{1}
Department of Electronic Engineering, Tsinghua University, Beijing, China \\
E-mail: hl17@mails.tsinghua.edu.cn, \{yangyy, sunjiasong\}@tsinghua.edu.cn}
\authorblockA{%
\authorrefmark{2}
University of Chinese Academy of Sciences, Beijing, China
}
\authorblockA{%
\authorrefmark{3}
Key Laboratory of Speech Acoustics and Content Understanding, Institute of Acoustics, Chinese Academy of Sciences, China \\
E-mail: \{chenggaofeng, zhangpengyuan\}@hccl.ioa.ac.cn}
\authorblockA{%
\authorrefmark{4}
North China Power Engineering CO., LTD. of China Power Engineering Consulting Group \\
Email: xushumin81@tom.com
}
}

\maketitle
\thispagestyle{empty}

\begin{abstract}

Utterance-level permutation invariant training (uPIT) has achieved promising progress on single-channel multi-talker speech separation task. 
Long short-term memory (LSTM) and bidirectional LSTM (BLSTM) are widely used as the separation networks of uPIT, i.e. uPIT-LSTM and uPIT-BLSTM. 
uPIT-LSTM has lower latency but worse performance, 
while uPIT-BLSTM has better performance but higher latency. 
In this paper, we propose using latency-controlled BLSTM (LC-BLSTM) during inference to fulfill low-latency and good-performance speech separation. 
To find a better training strategy for BLSTM-based separation network, 
chunk-level PIT (cPIT) and uPIT are compared. 
The experimental results show that uPIT outperforms cPIT when LC-BLSTM is used during inference. 
It is also found that the inter-chunk speaker tracing (ST) can further improve the separation performance of uPIT-LC-BLSTM. 
Evaluated on the WSJ0 two-talker mixed-speech separation task, 
the absolute gap of signal-to-distortion ratio (SDR) between uPIT-BLSTM and uPIT-LC-BLSTM is reduced to within 0.7 dB.
\end{abstract}

\section{Introduction}
\label{sec:introduction}

Many advancements have been observed for monaural multi-talker speech separation 
\cite{wang2006computational,schmidt2006single,hershey2016deep,isik2016single,chen2017deep,luo2018tasnet,yu2017permutation,kolbaek2017multitalker,luo2018real}, 
known as cocktail party problem \cite{haykin2005cocktail}, 
which is meaningful to many practical applications,
such as human-machine interaction, automatic meeting transcription etc. 
With the development of deep learning\cite{lecun2015deep}, 
a lot of innovations have been proposed, 
such as deep clustering \cite{hershey2016deep,isik2016single}, 
deep attractor network \cite{chen2017deep}, 
time-domain audio separation network \cite{luo2018tasnet,luo2018real} 
and permutation invariant training (PIT) \cite{yu2017permutation,kolbaek2017multitalker}.

Deep clustering \cite{hershey2016deep,isik2016single} projects the time-frequency (TF) units into an embedding space, 
with a clustering algorithm to generate a partition of TF units, 
which assumes that each bin belongs to only one speaker. 
However, the separation under the embedding space may be not the optimal technique.

Deep attractor network \cite{chen2017deep} also learns a high-dimensional representation of the mixed speech 
with some attractor points in the embedding space to 
attract all the TF units corresponding to the target speaker. 
However, the estimation of attractor points has a high computational cost.

PIT \cite{yu2017permutation} is an end-to-end speech separation method, 
which gives an elegant solution to the training label permutation problem
\cite{chen2017deep,yu2017permutation}.
It is extended to utterance-level PIT (uPIT) \cite{kolbaek2017multitalker}
with an utterance-level cost function to further improve the performance. 
Because uPIT is simple and well-performed, it draws a lot of attention
\cite{luo2018tasnet,luo2018real,yu2017recognizing,qian2018single,chen2018progressive,chang2018adaptive,tan2018knowledge,seki2018purely,chang2018end,xu2018single}. 
LSTM \cite{hochreiter1997long,gers2000learning,gers2002learning} 
and BLSTM \cite{graves2005framewise,graves2013speech} are widely used for 
uPIT to exploit utterance-level long time dependency.
Although uPIT-BLSTM outperforms uPIT-LSTM, 
its inference latency is as long as the utterance,
which hampers its applications in many scenarios.


To reduce the latency of BLSTM-based acoustic model
on automatic speech recognition (ASR) tasks, 
context-sensitive chunk (CSC) \cite{chen2016training}, 
which is the chunk with appended contextual frames, 
is proposed for both training and decoding.
In \cite{zhang2016highway}, CSC-BLSTM is extended to latency-controlled BLSTM (LC-BLSTM), 
which directly carries over the left contextual information from previous chunk of the same utterance
to reduce the computational cost and improve the recognition accuracy.

In this paper, inspired by LC-BLSTM-based acoustic model on ASR tasks,
uPIT-LC-BLSTM for low-latency speech separation is proposed,
which splits an utterance into non-overlapping chunks with future contextual frames 
during inference to reduce the latency from utterance-level to chunk-level.
The chunk-level PIT (cPIT) of BLSTM is also proposed, 
but the preliminary experiments indicate that cPIT is inferior to uPIT. 
uPIT-LC-BLSTM propagates BLSTM's forward hidden states across chunks, 
which helps keep the speaker consistency across chunks. 
Meanwhile, an inter-chunk speaker tracing (ST) algorithm is proposed
to further improve the performance of uPIT-LC-BLSTM. 
Experiments evaluated on the WSJ0 two-talker mixed-speech separation task
show that uPIT-LC-BLSTM with ST only loses a little when compared to uPIT-BLSTM.

The paper starts by briefly describing  prior work in Section \ref{sec:prior}.
The cPIT, uPIT-LC-BLSTM and speaker tracing algorithm 
are described in Section \ref{sec:cpit}.
The experimental setup and results are discussed in Section \ref{sec:exp}. 
Section \ref{sec:con} presents the conclusions.

\section{Prior Work}
\label{sec:prior}
\subsection{Monaural Speech Separation}
The goal of single-channel multi-talker speech separation is to separate the individual source signals from the mixed audio. 
Let us denote $S$ source signals as $\textbf{x}_s(t), s=1,...,S$ 
and the microphone receives mixed audio $\textbf{y}(t)=\sum_{s=1}^{S}\textbf{x}_s(t)$.
The separation is often carried out in the time-frequency (TF) domain, 
where the task is to reconstruct the short-time Fourier transform (STFT) of each individual source signal.  
The STFT of the mixed signal is $\textbf{Y}(t,f)=\sum_{s=1}^{S}\textbf{X}_s(t,f)$, 
where $\textbf{Y}(t,f)$ is the TF unit at frame $t$ and frequency $f$.

The STFT reconstruction of each source can be done 
by estimating $S$ masks $\hat{\textbf{M}}_s(t,f),s=1,...,S$. 
We use phase sensitive mask (PSM) here:
$\textbf{M}_s(t,f)=\frac{|\textbf{X}_s(t,f)|}{|\textbf{Y}(t,f)|}\cos(\theta_\textbf{Y}(t,f)-\theta_{\textbf{X}_s}(t,f))$, 
where $|\textbf{Y}|$ and $\theta_\textbf{Y}$ are the magnitude and phase of $\textbf{Y}$ respectively. 
With an estimated mask $\hat{\textbf{M}}_s(t,f)$ and the mixed STFT,
the STFT of source $s$ is $\hat{\textbf{X}}_s(t,f)=\hat{\mathbf{M}}_s(t,f)\cdot |\mathbf{Y}(t,f)|\cdot e^{j \theta_{\textbf{Y}}(t,f)}$,
where $j$ is imaginary unit.

The straightforward mask-based separation methods 
based on deep learning are to use neural network to 
estimate masks for $S$ source signals and then
minimize the mean square error (MSE) between estimated and target magnitudes.
For PSM, the cost function is as follows:
\begin{equation}
\mathcal{J}_{psm}=\frac{1}{B}
\sum_{s=1}^{S}||\hat{\mathbf{M}}_s\circ |\mathbf{Y}|-|\textbf{X}|_s\circ \cos(\theta_\textbf{Y}-\theta_{\textbf{X}_s})||^2_F
\end{equation}
where $B=T\times F\times S$ is the total number of TF units, 
$\circ$ is the element-wise product and ${||\cdot||}_F$ is the Frobenius norm. 

\begin{figure}[t]
	\centering
	\includegraphics[width=2.6in,height=2.85in]{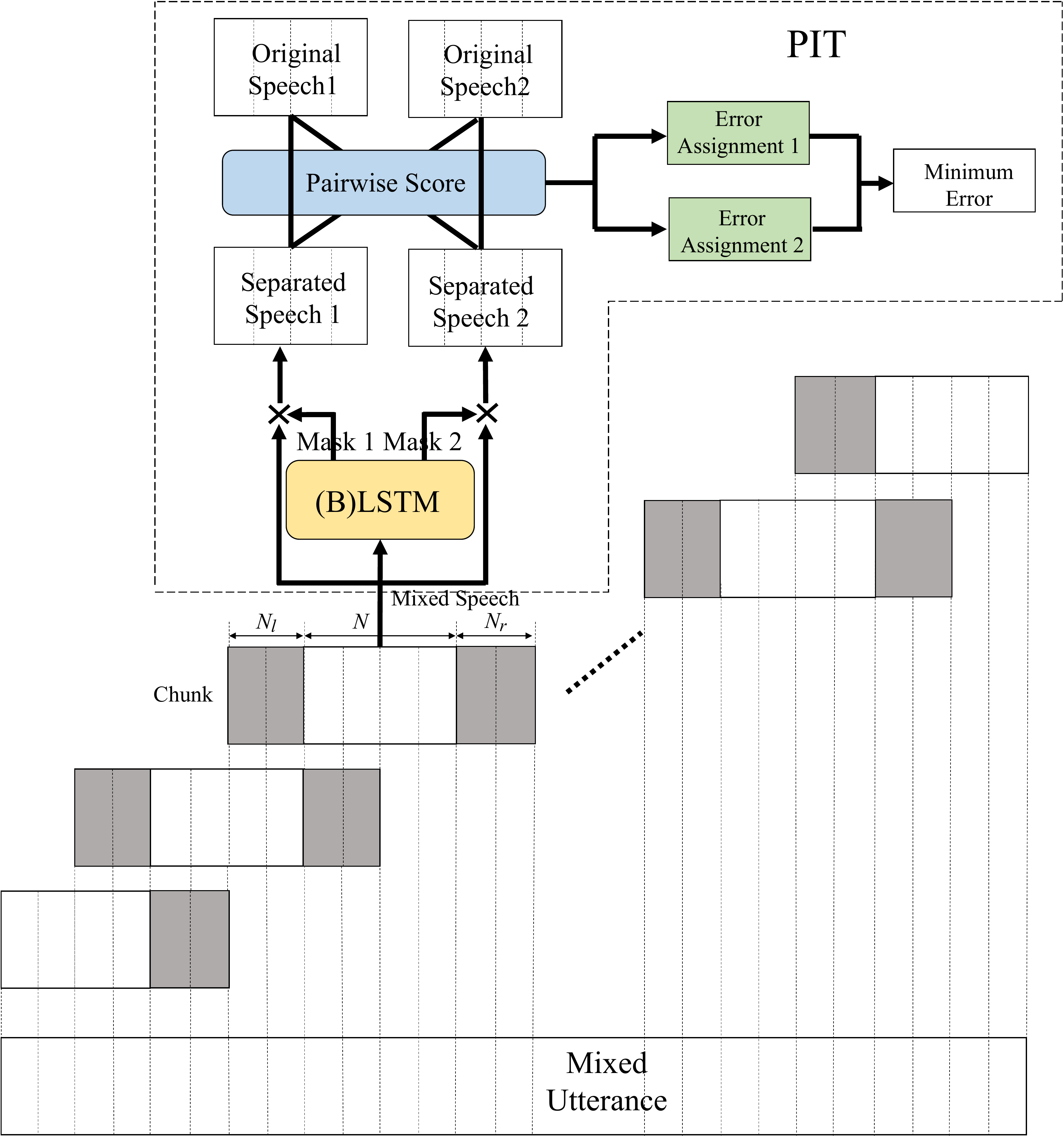}
	\caption{The architecture of cPIT, whose main idea is to split an utterance into chunks.
			The main chunk has $N$ frames, 
			with appended $N_l$ left and $N_r$ right contextual frames.
			For the first/last chunk of each utterance, 
			no left/right contextual frames are appended.
			The appended frames are only used to provide context information and
			do not generate error signals during training.
			LC-BLSTM does not need left contextual frames.}
	\label{fig:cpit}
\end{figure}

\subsection{Utterance-level Permutation Invariant Training}
The cost function mentioned above is a good way for some simple cases. 
For example, when a priori convention can be learned,
we can force the speakers with higher energy (or male speakers) to be the first output, 
and those with lower energy (or female speakers) to be the second output.
However, when the energy difference is small or two speakers have the same gender,
a problem named label permutation \cite{chen2017deep,yu2017permutation} is introduced,
where the permutation of two output streams is unknown.

PIT \cite{yu2017permutation} has eliminated the label permutation problem, 
while it faces another problem named speaker tracing, 
which is solved by extending PIT with an utterance-level cost function, i.e. uPIT \cite{kolbaek2017multitalker}, 
to force the separation of the same speaker into the same output stream.
The cost function of uPIT is as follows:
\begin{equation}
\mathcal{J}=\frac{1}{B}
\sum_{s=1}^{S}||\hat{\mathbf{M}}_s\circ |\mathbf{Y}|-|\textbf{X}|_{\phi^*(s)}\circ \cos(\theta_\textbf{Y}-\theta_{\textbf{X}_{\phi^*(s)}})||^2_F
\end{equation}
where $\phi^*$ is the permutation that minimizes the separation error:
\begin{equation}
\phi^*=\arg \min_{\phi \in \mathcal{P}} 
\sum_{s=1}^{S}||\hat{\mathbf{M}}_s\circ |\mathbf{Y}|-|\textbf{X}|_{\phi}\circ \cos(\theta_\textbf{Y}-\theta_{\textbf{X}_{\phi}})||^2_F
\end{equation}
where $\mathcal{P}$ is the set of all $S!$ permutations. 
As illustrated in the area surrounded by dotted lines in Figure \ref{fig:cpit}, 
PIT computes MSE between estimated and target magnitudes using all possible permutations, 
and the minimum error is used for back propagation. 

\subsection{CSC-BLSTM and LC-BLSTM}
BLSTM is often used in uPIT-based speech separation systems 
for its capacity of modeling long time dependency in forward and backward directions 
\cite{kolbaek2017multitalker,qian2018single,yu2017recognizing,chen2018progressive,chang2018adaptive,tan2018knowledge,seki2018purely,chang2018end}.
BLSTM has a high latency as long as the utterance.
Since BLSTM is one of the state-of-the-art acoustic models on ASR tasks
\cite{graves2005framewise,graves2013speech,chen2016training,zhang2016highway,xue2017improving,xiong2018microsoft,xiong2016achieving,cheng2018output,wenjie2018investigation,Han2018Densely},
there have been some relative works to address the latency problem \cite{chen2016training,zhang2016highway,peddinti2018low}. 

In \cite{chen2016training}, 
context-sensitive chunk (CSC) with left and right contextual frames 
to initialize the forward and backward LSTM
is used for both training and decoding, 
which reduces the decoding latency from utterance-level to chunk-level. 
CSC-BLSTM is extended to LC-BLSTM by directly carrying over the
left contextual information from previous chunk of the same utterance into current chunk \cite{zhang2016highway}, 
where the latency can be determined by the number of right contextual frames 
and modified by users to get a trade-off between performance and latency.

\section{Proposed Methods}
\label{sec:cpit}

\subsection{Chunk-level PIT}
As illustrated in Figure \ref{fig:cpit}, 
the proposed cPIT splits an utterance into context-sensitive chunks, 
where main chunks (without contextual frames) do not overlap.
Since the lengths of chunks are very close 
(no longer than $N_l+N+N_r$),
we do not need to do zero padding frequently during training,
so the training can be sped up significantly when compared to uPIT.
Besides, we evaluate whether cPIT is beneficial for chunk-level inference.

\subsection{cPIT-LC-BLSTM and uPIT-LC-BLSTM}
Inference can also be done at the utterance level or chunk level.
If we simply infer at the chunk level, i.e. use CSC-BLSTM, 
the output streams of main chunks in the same utterance 
are spliced to compose utterance-level separated results.
However, permutation may change across neighboring chunks. 
For instance, in two-speaker case, the output permutation may be 1-1 
(the first output stream corresponds to the first speaker) and 2-2 in previous chunk, 
while it may change to 1-2 (the first output stream corresponds to the second speaker) 
and 2-1 in current chunk. 
If the output streams of these two chunks are simply spliced,
the separated speech may face the speaker inconsistency problem.

The first proposed method to alleviate the problem is to replace CSC-BLSTM with LC-BLSTM.
The only difference between them is that
LC-BLSTM copies the forward hidden states from previous chunk directly
and does not need left contextual frames,
while CSC-BLSTM uses left contextual frames to initialize forward LSTM.
They both need right contextual frames to initialize backward LSTM.
There are two advantages in using LC-BLSTM.
Firstly, computational cost is reduced by $\frac{N_l}{N_l+N+N_r}$ with the
removing the left initialization operation.
Secondly, it helps keep the forward hidden states continuous
across neighboring chunks, which is beneficial for modeling a broader left context
and to some extent alleviates the speaker inconsistency problem.

With the model trained at the chunk level or utterance level, 
cPIT-LC-BLSTM or uPIT-LC-BLSTM method is obtained.
Besides, some other denotations are also listed in Table \ref{naming}.

\begin{table}
	\scriptsize
	\centering\caption{For simplicity and clarity, 
		some denotations are listed.}
	\centering\begin{tabular}{|l|c|c|l|}
		\hline
		Denotation	 	&Model	&Training Strategy 	&Inferring Method 	\\ \hline
		uPIT-LSTM		&LSTM	&utterance-level PIT&utterance-level  	\\ \hline
		uPIT-BLSTM 		&~		&~					&utterance-level 	\\ 
		uPIT-CSC-BLSTM	&BLSTM	&utterance-level PIT&chunk-level (CSC)	\\ 
		uPIT-LC-BLSTM 	&~		&~					&chunk-level (LC)	\\ \hline
		cPIT-BLSTM 		&~		&~					&utterance-level	\\
		cPIT-CSC-BLSTM 	&BLSTM	&chunk-level PIT	&chunk-level (CSC)	\\
		cPIT-LC-BLSTM 	&~		&~					&chunk-level (LC)	\\
		\hline
	\end{tabular}
	\label{naming}
\end{table}

\subsection{Inter-chunk Speaker Tracing}
In \cite{yu2017permutation}, there is a huge performance gap 
between default assign (without ST) 
and optimal assign (assuming that all speakers are correctly traced), 
which can be reduced with ST algorithms.

In this paper, a simple ST algorithm is adopted to exploit the overlapping frames 
between two neighboring chunks.
Let us denote $\mathbf{O}_{t-1}^1$ and $\mathbf{O}_{t-1}^2$ as two output streams of overlapping frames in previous chunk,
and $\mathbf{O}_{t}^1$ and $\mathbf{O}_{t}^2$ as those in current chunk.
We compute pairwise MSE as PIT does:
\begin{equation}
E_1=\text{MSE}(\mathbf{O}_{t-1}^1,\mathbf{O}_{t}^1)+\text{MSE}(\mathbf{O}_{t-1}^2,\mathbf{O}_{t}^2)
\end{equation}
\begin{equation}
E_2=\text{MSE}(\mathbf{O}_{t-1}^1,\mathbf{O}_{t}^2)+\text{MSE}(\mathbf{O}_{t-1}^2,\mathbf{O}_{t}^1)
\end{equation}
If $E_1>\alpha E_2$, we consider there exists a change of output permutation, 
where $\alpha$ is the penalty factor and set to 2.0 by default.
There are two reasons to set $\alpha$ to 2.0 instead of 1.0. 
Firstly, we believe that the probability of permutation changing is smaller than that of the same permutation,
especially when LC-BLSTM is used. 
Secondly, more robustness is added into the system.
For example, if both speakers are silent in the overlapping frames, 
the two output streams are almost similar, 
and then setting $\alpha$ to 1.0 may lead to a false detection of permutation changing.

\begin{table}
	\scriptsize
	\centering\caption{SDR improvements (dB) for original mixtures and uPIT-(B)LSTM baselines. M/F stands for male/female.}
	\centering\begin{tabular}{|l|c|ccc|}
		\hline
		PIT Model &Average &M-F&F-F&M-M\\ \hline
		Mixtures  & 0.06	&0.06	&0.07	&0.06 \\ 
		\hline
		uPIT-LSTM \cite{kolbaek2017multitalker}& 7.0 &- &- &-\\
		uPIT-BLSTM \cite{kolbaek2017multitalker}& 9.4 &- &- &- \\
		\hline
		Our uPIT-LSTM & 7.16	&9.02	&3.80	&5.77 \\
		Our uPIT-BLSTM& 9.46	&10.90	&7.61	&8.11 \\ 
		\hline
	\end{tabular}
	\label{pit_baseline}
\end{table}

\section{Experiments and Results}
\label{sec:exp}
\subsection{Experimental Setup}
The dataset is the same as the two-talker mixed dataset in \cite{hershey2016deep,isik2016single,luo2018tasnet,yu2017permutation,kolbaek2017multitalker}, 
except that the sample rate is 16 kHz.
It is generated by mixing the utterances in WSJ0 corpus at various signal-to-noise ratios uniformly chosen between 0 dB and 5 dB, 
and has 20k, 5k and 3k mixtures for training, validation and testing respectively. 
The 30-hour training set and 10-hour validation set are generated from \texttt{si\_tr\_s} using 49 male and 51 female speakers.
The 5-hour testing set is generated from \texttt{si\_dt\_05} and \texttt{si\_et\_05} using 16 speakers.

The input to the model is the magnitude of mixture's STFT,
which is extracted with a frame size 32 ms and 16 ms shift, \
and has 257 frequency sub-band.
The PIT model has a fully-connected layer, 
3 (B)LSTM layers and two output layers.
The dimension of LSTM cell is 640, so each BLSTM layer has 1280 units.
We use ReLU \cite{nair2010rectified} as the activation function of two output layers,
and two output masks have the same dimension as that of input.
The input mixed magnitude is multiplied by two masks respectively 
to get two separated magnitudes, 
and then use the phase of mixed speech and inverse STFT to get the separated audios. 
Signal-to-distortion ratio (SDR) \cite{vincent2006performance} is used to evaluate the performance of separation.

Tensorflow \cite{abadi2016tensorflow} is used to build the systems. 
The validation set is only used for tuning the learning rate as it will be halved by 0.7 when the loss on validation set increases. 
The initial learning rate is 0.0005.
Dropout is applied to BLSTM layers with a rate 0.5.
For faster evaluation, all models are trained for 32 epochs.
When training at the utterance level, each minibatch contains 10 random utterances.
When training at the chunk level, each minibatch contains 100 random chunks.

\subsection{uPIT Baselines}

Table \ref{pit_baseline} presents the SDR improvements of baseline uPIT-(B)LSTM. 
It is obvious that uPIT-BLSTM is far better than uPIT-LSTM. 
It is also noticed that the same-gender separation is more difficult, especially female-female separation.
Although the size of our model is smaller than that in \cite{kolbaek2017multitalker} and 
we trained for fewer epochs, the obtained results are 
comparable with the baseline results in \cite{kolbaek2017multitalker}.

\subsection{cPIT v.s. uPIT}

\begin{table}
	\scriptsize
	\centering\caption{Average SDR improvements (dB) for BLSTM trained with cPIT or uPIT.
		 Speaker tracing (ST) is used to improve the performance of CSC-BLSTM and LC-BLSTM.
		 The absolute gap (Abs. Gap) is compared to uPIT-BLSTM.}
	\centering\begin{tabular}{|l|cc|}
		\hline
		 Method 					&  SDR  		& Abs. Gap\\ \hline
		cPIT-CSC-BLSTM 				& 8.00 			& -1.46 \\
		cPIT-CSC-BLSTM + ST			& {8.72} 		& -0.74 \\
		cPIT-LC-BLSTM 				& 8.61 			& -0.85 \\
		cPIT-LC-BLSTM + ST			& 8.71 			& -0.75 \\
		\hline
		cPIT-BLSTM					& 8.73 			& -0.73 \\
		\hline
		uPIT-CSC-BLSTM				& 8.09 			& -1.37 \\
		uPIT-CSC-BLSTM + ST	   		& 9.10 			& -0.36 \\
		uPIT-LC-BLSTM 			   	& 8.98 			& -0.48 \\
		uPIT-LC-BLSTM + ST	    	& 9.16 			& -0.30 \\
		\hline
		uPIT-BLSTM					& 9.46 			& -\\
		\hline
	\end{tabular}
	\label{blstm_cpit1}
\end{table}

\begin{figure*}[t]
	\centering
	\includegraphics[scale=0.405]{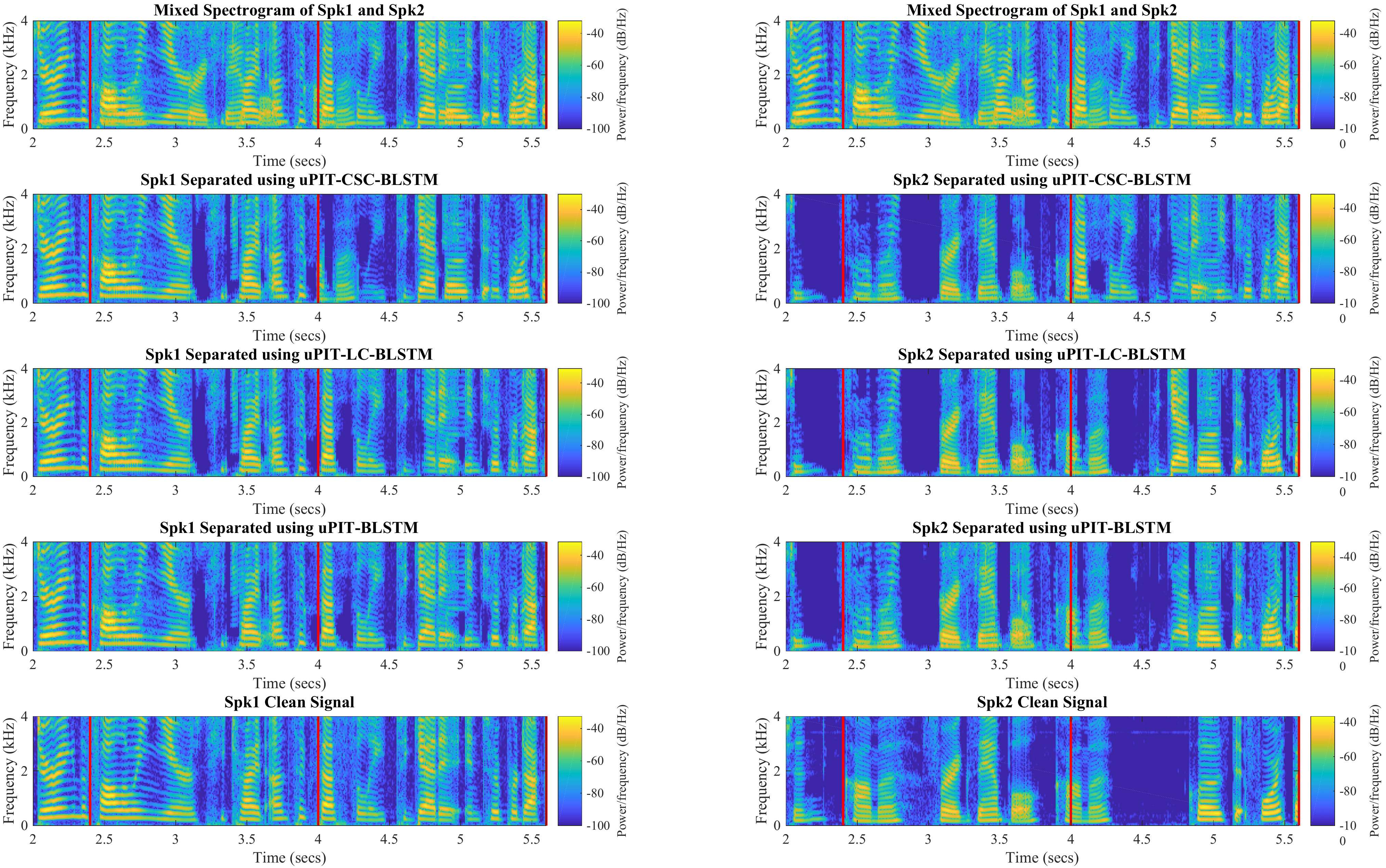}
	\caption{Permutation changing problem is alleviated by LC-BLSTM.
		The mixed/clean spectrograms of two speakers are shown in the first/last row respectively.
		The second, third and last rows are the separated spectrograms
		using uPIT-CSC-BLSTM, uPIT-LC-BLSTM and uPIT-BLSTM respectively.
		The vertical red lines are the borders of chunks.
		In the second row, the speakers exchanges in the last chunk when using CSC-BLSTM,
		and it is solved in the third row when using LC-BLSTM.}
	\label{fig:result}
\end{figure*}

As described before, the model for inference can be trained at the utterance level or chunk level.
We trained one BLSTM at the chunk level with $N_l=50, N=100, N_r=50$, 
and compared it with the baseline BLSTM trained at the utterance level.
We present the SDR results in Table \ref{blstm_cpit1}.
Here, we consider four inferring methods: cPIT-CSC-BLSTM, uPIT-CSC-BLSTM, cPIT-BLSTM and uPIT-BLSTM.
Generally, the model trained at the utterance level performs better.



\subsection{CSC-BLSTM v.s. LC-BLSTM}
Here we compare two inferring methods: CSC-BLSTM and LC-BLSTM.
As illustrated in Table \ref{blstm_cpit1}, LC-BLSTM outperforms CSC-BLSTM significantly,
with improvements of 0.61 dB when using the model trained at the chunk level
and 0.89 dB when using the model trained at the utterance level.
Besides, uPIT-LC-BLSTM outperforms cPIT-LC-BLSTM significantly.

To prove LC-BLSTM helps alleviate the speaker inconsistency problem, 
an example is shown in Figure \ref{fig:result}.
As illustrated, 
there exists a change of permutation in the last chunk when using uPIT-CSC-BLSTM.
Also, the spectrograms separated by uPIT-LC-BLSTM and uPIT-BLSTM are quite similar.

\subsection{Inter-chunk Speaker Tracing}
As illustrated in Table \ref{blstm_cpit1}, 
ST can further improve the performance of both CSC-BLSTM and LC-BLSTM.
For the model trained at the chunk level,
ST improves the cPIT-CSC-BLSTM and cPIT-LC-BLSTM by 0.72 dB and 0.1 dB respectively.
For the model trained at the utterance level,
ST improves the uPIT-CSC-BLSTM and uPIT-LC-BLSTM by 1.01 dB and 0.18 dB respectively,
where the improvements are more obvious.

Finally, uPIT-LC-BLSTM with ST achieves the best results of chunk-level inference,
which is slightly worse than that of uPIT-BLSTM with a gap 0.3 dB, 
but is significantly better than that of uPIT-LSTM with a gain of 2.0 dB.

\subsection{Trade-off between Latency and Performance}
\begin{table}
	\scriptsize
	\centering\caption{Average SDR improvements (dB) and latency (defined as $16\times N_r$ ms as that in \cite{zhang2016highway}) for uPIT-LC-BLSTM.}
	\centering\begin{tabular}{|l|c|cc|c|}
		\hline
		Method&$N_r$ & SDR & Abs. Gap & Latency (ms)\\ \hline
		uPIT-LC-BLSTM&0  & {8.76} & -0.70 & 0 \\
		\hline
		& 10  &{8.81} & -0.55 & 160\\
		& 25  &{9.02} & -0.44 & 400\\
		uPIT-LC-BLSTM + ST & 35  &{9.07} & -0.39 & 560\\
		& 50 &{9.16} & -0.30 & 800 \\
		& 100 &{9.26}  & -0.20 & 1600\\
		\hline
		\multicolumn{2}{|l|}{uPIT-BLSTM}& 9.46  & - & utterance-level\\ \hline
		\multicolumn{2}{|l|}{uPIT-LSTM} & 7.16  & -2.30 & 0 \\
		\hline
	\end{tabular}
	\label{blstm_cpit2}
\end{table}

The latency of above chunk configuration is $50\times 16$ ms $=800$ ms
(defined as $16\times N_r$ ms as that in \cite{zhang2016highway}), 
which is quite high for low-latency applications. 
Here, we keep $N_l$ and $N$ fixed (Note $N_l$ is useless for LC-BLSTM), 
and change the value of $N_r$ to evaluate the performance with different latency, 
and the results are illustrated in Table \ref{blstm_cpit2}.

Generally, SDR decreases as $N_r$ decreases. 
Note that when $N_r=0$, we cannot perform ST for LC-BLSTM,
since there is no overlapping frame.
Even though  $N_r$ is $0$, 
uPIT-LC-BLSTM still outperforms uPIT-LSTM with a gain of 1.6 dB,
and has a gap of 0.7 dB when compared to uPIT-BLSTM.

\section{Conclusions}
\label{sec:con}
In this paper, we explored uPIT-LC-BLSTM on single-channel multi-talker speech separation task
to reduce the latency of uPIT-BLSTM from utterance-level to chunk-level.
To reduce the SDR gap between uPIT-LC-BLSTM and uPIT-BLSTM, 
inter-chunk speaker tracing was proposed to further
alleviate the permutation changing problem across neighboring chunks.
Besides, a trade-off between inference latency and separation performance could be obtained 
according to the actual demand by setting the number of right contextual frames.
In the future, we plan to combine the uPIT-LC-BLSTM with cross entropy for
directly multi-talker speech recognition \cite{yu2017recognizing,qian2018single,chen2018progressive,chang2018adaptive,tan2018knowledge}.

\section*{Acknowledgements}
This work is partially supported by the National Natural Science Foundation of China (Nos. 11590774, 11590770). 

\bibliographystyle{IEEEtran}

\bibliography{mybib}

\end{document}